**Built-in surface field effect on the 180-degree domain wall structure in ferroics**

Eugene A. Eliseev,[*]

Institute for Problems of Materials Science, NAS of Ukraine,

Krjijanovskogo 3, 03142 Kiev, Ukraine,

We consider the influence of the built-in surface field on the 180-degree domain wall profile in primary ferroics within Landau-Ginsburg-Devonshire phenomenological approach. We predict the effect of domain wall bending near the surface caused by the built-in field and derived corresponding approximate analytical expressions. At that the higher is the surface energy contribution (i.e. the smaller is corresponding extrapolation length) and/or the higher is the field, the stronger is the bending effect. Built-in surface field is one of the possible mechanisms of domain wall near surface broadening recently observed in the ferroics.

Extra broad domain walls with width about 10-100 nm were observed at $LiNbO_3$ surface by Piezoelectric Force Microscopy (PFM).[1] Also recent experimental results of scanning nonlinear dielectric microscopy (SNDM) [2, 3] reveal that wall width near the surface is several times (up to five) higher than in the bulk of thin films of $LiTaO_3$. Recently the effect of surface wall broadening in ferroelectrics was attributed to the formation of double electric layer break at the wall-surface junction.[4, 5] However, this mechanism is specific to the domain walls in ferroelectrics with polarization having normal to the surface component.

For the case of ferroelastic the surface effect on elastic twins in proper ferroelastic were studied in details by Salje et al.[6, 7, 8] It was found[7] that nonzero internal stress exists at a domain wall due to coupling between primary order parameter (shear strain) and dilatation strains.

---

[*] E-mail: eliseev@i.com.ua

Relaxation of stress normal components at the free surface led to either the wall widening or narrowing near the surface depending on the sign of surface curvature.

Rychetsky[9] considered the internal stresses occurring around domain walls and showed that the stress can cause deformation of the crystal surface. The surface distortions have been calculated for the case of anti-phase boundaries in tetragonal crystals.

The message of the paper is to consider the influence of the built-in surface field on the anti-parallel 180°-domain wall profile in primary ferroics within Landau-Ginsburg-Devonshire (LGD) phenomenological approach [10, 11, 12]. Built-in surface field causes phase transition smearing in thin films[13, 14] and stimulates domains nucleation during polarization reversal.[15] This field could be related to either spontaneous symmetry breaking[14, 15] or misfit strain between film and substrate[13] via surface piezoelectric effect.[16, 17]

For ferroelectric and ferromagnetic with order parameter normal to the surface depolarization and demagnetization fields play important role on the structure of the domain walls, as it was shown recently for ferroelectrics taking into account incomplete screening due to the extrapolation length effect.[4] Note, that for ferromagnetic films the demagnetization field is known to cause the Bloch type domain wall transformation into a Neel one with thickness decrease.[18] To study the influence of the surface effects on the domain wall itself, we choose the problem geometry for ferroelectric or ferromagnetic *without depolarization* or *demagnetization fields*, which is possible e.g. for in-plain components of order parameter, depending only on coordinates normal to the order parameter (so that divergence of the order parameter vector is zero, $\mathrm{div}(\mathbf{\eta}) = 0$, and sources of depolarization field is absent). Note that for antiferromagnetics or antiferroelectrics the inhomogeneity of order parameters do not produce any conjugated field, causing decrease of order parameter.

Geometry of the problem is presented in Fig.1. In this case polarization (magnetization) pointed along y-axis do not produce depolarization (demagnetization) field.

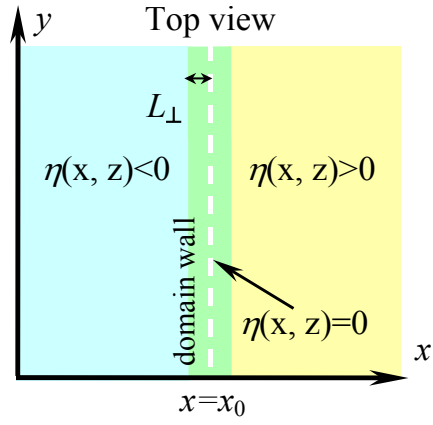

**Fig. 1.** 180°-domain wall near the sample surface $z=0$. Order parameter $\eta(x,z)$ has different signs in different domains. Bulk domain wall profile is $x=x_0$.

For the case of primary ferroics LGD expansion of bulk Gibbs free energy on the order parameter $\eta$ powers (components of polarization, magnetization or strain tensor for ferroelectric, ferromagnetic or ferroelastic media respectively):

$$G_V = \int_V d^3r \left( \frac{\alpha}{2}\eta^2 + \frac{\beta}{4}\eta^4 + \frac{\gamma}{2}\left(\frac{\partial \eta}{\partial z}\right)^2 + \frac{\xi}{2}\left(\left(\frac{\partial \eta}{\partial y}\right)^2 + \left(\frac{\partial \eta}{\partial x}\right)^2\right) \right). \tag{1a}$$

Hereinafter we neglect the coupling with elastic stresses (i.e. magnetostriction and electrostriction terms). Coefficient $\alpha \sim (T-T_C)$ linearly depends on the temperature $T$, so that it is negative below Curie temperature $T_C$. Constants $\gamma$ and $\xi$ determine the strength of gradient energy. Coefficients of the energy expansion $\beta$, $\gamma$ and $\xi$ are positive, they may also weakly depend on temperature.

In the phenomenological theory framework the surface effects should be taken into account by introducing the surface energy, depending on the order parameter as follows:

$$G_S = \int_S d^2r \left( \frac{\alpha_S}{2}\eta^2 - \sigma\eta \right). \tag{1b}$$

Expansion coefficient $\alpha_S$ is assumed positive (otherwise higher expansion terms should be considered). Built-in surface field $\sigma$ may be caused by different physical mechanisms for ferroelastic, ferroelectric and antiferromagnetic media.

For instance, $\sigma$ is the component the intrinsic surface stress tensor for ferroelastic with order parameter as the component of strain tensor $u_{ij}$. Namely, for this case surface energy can be rewritten as $\int_S \left( \alpha^S_{ijkl} u_{ij} u_{kl}/2 + \xi_{ij} u_{ij} \right) d^2 r$. Here summation on repeated indices are assumed, $\xi_{ij}$ is the intrinsic surface stress tensor.[19, 20, 21] It should be noted that for flat surfaces only in-plane components of $\xi_{ij}$ would be invariant under in-plane rotations. Thus, only in-plane components of stress could be induced by surface stress effect.

For the ferroelectrics with order parameter $\eta$ as the component of polarization vector $P_i$, $\sigma$ originates from the symmetry breaking near the surface. For instance, in the case of polarization, normal to the film surface, vanishing of inversion center leads to the appearance of $\sigma$ (see e.g. Refs. [13, 14]). In this case one can write surface energy as $\int_S \left( \alpha^S_{ij} P_i P_j /2 + b P_l n_l + \mu_{ijkl} u_{ij} P_k n_l \right) d^2 r$. Here $n_i$ is the components of the normal to the surface, the second term is responsible for surface local electric field, oriented perpendicular to the surface.[13, 14] The latter term is related to the surface flexoelectric effect.[16, 17] Using the symmetry of bulk flexoelectric effect for cubic materials,[22, 23] it is easy to estimate that these terms couple in-plane polarization components to shear strains. Although these shear strains are hardly to be induced by surface stresses, they may appear due to the reconstruction of surface[24] or the influence of interface material.

It should be noted, that for the case of ferromagnetic or superconductor order parameter built-in surface field, coupled to corresponding order parameter, could not be considered, since surface could eliminate the symmetry elements corresponding to spatial symmetry but it cannot eliminate the time-reversal or gauge transformations which change ferromagnetic and superconducting order parameters.[14] However, the situation with antiferromagnetic media is more complex, since the allowed symmetry groups for antiferromagnetic may not include time-reversal operation and thus these media may possess piezomagnetic effect[10] as well as terms in surface energy linear on order parameter.

One can conclude from the above consideration, that for many different type ferroics media appearance of linear on order parameter terms in the surface energy is quite possible. Here, we consider the impact of these terms on the surface structure of ferroic domain walls.

For the second order phase transitions considered hereinafter, LGD equation for the order parameter $\eta(x,z)$ has the form:

$$\alpha\eta + \beta\eta^3 - \gamma\frac{\partial^2\eta}{\partial z^2} - \xi\left(\frac{\partial^2\eta}{\partial x^2} + \frac{\partial^2\eta}{\partial y^2}\right) = 0 \ . \tag{2a}$$

Further let us consider the case of semi-infinite sample with surface $z=0$. Inhomogeneous boundary condition for the order parameter is:

$$\left(\eta - \lambda\frac{\partial\eta}{\partial z}\right)\bigg|_{z=0} = \eta_t(x). \tag{2b}$$

Here $\lambda = \gamma/\alpha_S$ is the extrapolation length (see e.g. Refs [25], [26], [27]), built-in surface order parameter is directly proportional to the surface field as $\eta_t(x,y) = \sigma(x,y)/\alpha_S$.

Bulk 1D-solution for 180°-domain wall is $\eta_0(x) = \eta_S \tanh((x-x_0)/2L_\perp)$, where correlation length $L_\perp = \sqrt{-\xi/2\alpha}$ and spontaneous value $\eta_S^2 = -\alpha/\beta$. It is no more a solution of system (2) for the case $|\lambda| < \infty$.

For the typical case $|\eta_t| < \eta_S$ linearized solution of Eqs.(2) can be found as $\eta(x,z) = \eta_S \tanh\left(\frac{x-x_0}{2L_\perp}\right) + p(x,z)$, where the perturbation $p(x,z)$ satisfies the inhomogeneous boundary problem:

$$\left(-2\alpha + 3\alpha\,\text{sech}^2\left(\frac{x-x_0}{2L_\perp}\right)\right)p - \gamma\frac{\partial^2 p}{\partial z^2} - \xi\frac{\partial^2 p}{\partial x^2} = 0, \tag{3a}$$

$$\left(p - \lambda \frac{\partial p}{\partial z}\right)\bigg|_{z=0} = \eta_t(x) - \eta_S \tanh\left(\frac{x - x_0}{2L_\perp}\right). \tag{3b}$$

Where the hyperbolic function definition $\operatorname{sech}^2(x) = 1 - \tanh^2(x)$ was used.

Looking for the solution of Eqs.(3) in the form $p(x,z) = \int_{k \geq 0} dk\, q(k,x) \exp(-kz)$, one obtains the equations for the spectrum $q(k,x)$:

$$\left(-2\alpha + 3\alpha \operatorname{sech}^2\left(\frac{x - x_0}{2L_\perp}\right) + \gamma k^2\right) q(k,x) - \xi \frac{d^2}{dx^2} q(k,x) = 0, \tag{4}$$

Eq.(4) is linear inhomogeneous equation with x-dependent coefficient. The solution of homogeneous equation for $q(k,x)$ was derived as proposed in Ref.[28], namely:

$$q(k,x) = A(k) f(k, x - x_0) + B(k) f(k, x_0 - x), \tag{5}$$

$$f(k,x) = \exp\left(i\kappa(k) \frac{x}{2L_\perp}\right)\left(\kappa^2(k) - 2 + 3\operatorname{sech}^2\left(\frac{x}{2L_\perp}\right) + 3i\kappa(k) \tanh\left(\frac{x}{2L_\perp}\right)\right). \tag{6}$$

Then one obtains integral equation for $A(k)$ and $B(k)$ from the boundary conditions (3b). For particular case $\eta_t = \text{const}$, approximate analytical solution was derived as (see supplementary materials):

$$\eta(x,z) \approx \eta_S \left(1 - \frac{\exp(-z/L_z)}{1 + \lambda/L_z}\right) \tanh\left(\frac{x - x_0}{2L_\perp}\right) + \eta_t \frac{\exp(-z/L_z)}{1 + \lambda/L_z} \tag{7}$$

By definition $L_\perp = \sqrt{-\xi/2\alpha}$, $L_z = \sqrt{-\gamma/2\alpha}$. First term in Eq. (7) is the bulk solution, but with amplitude depending on the depth due to the reduce of order parameter on surface (extrapolation length effect), the second term is due to the built-in surface field. Solution (7) reproduces both bulk domain wall far from surface ($z \to \infty$) and qualitative behavior of order parameter at the surface far from domain wall (and even quantitative at high values of $\lambda$).

Domain wall position corresponding to Eq. (7) is given by following expression:

$$x(z) \approx x_0 - 2L_\perp \operatorname{arctanh}\left(\frac{\eta_t}{\eta_S} \frac{\exp(-z/L_z)}{1 + \lambda/L_z - \exp(-z/L_z)}\right). \tag{8}$$

For the case $|\eta_t| < \eta_S \lambda/L_z$, the wall reaches the surface $z=0$, while for $|\eta_t| > \eta_S \lambda/L_z$ the built-in surface field induces ordered surface state instead of the wall bending. Eqs.(7)-(8) describe the effect of 180°-domain wall bending near the surface as demonstrated in Figs.2.

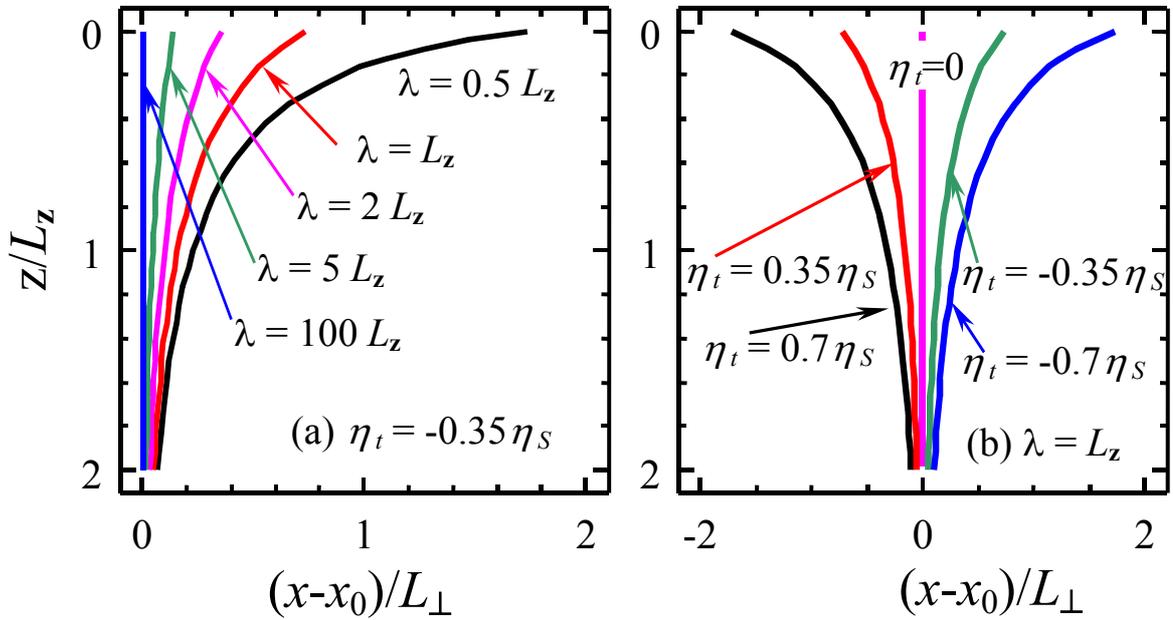

**Fig. 2**. (Color online) Domain wall profiles (curves $\eta(x,z)=0$) (a) for different extrapolation length $\lambda$ values (marked near the curves in $L_z$ units) and built-in field $\eta_t=-0.35\eta_S$; (b) for different ratio $\eta_t/\eta_S$ values (0, ±0.35, ±0.7) and $\lambda=L_z$. Depending on the sign, built-in field tends to re-polarize the "left" or "right" surface region and thus bend the domain wall. Bulk ($z \gg L_z$) domain wall profile is $x=x_0$.

Note that surface state formation in the field of defect were described earlier[29] for ferroelectric media.

The built-in surface field not only bends the wall, but also leads to changes in domain wall structure as whole. As one can see from Figs. 3a and 3c, the structure of the wall becomes asymmetric near the surface, and at sufficiently high values of build-in field domain wall could not reach the surface. Moreover, the domain wall width also undergoes changes due to surface field. It should be noted that the width of the domain wall is usually determined as the distance between the points, where the order parameter reaches the definite fraction of order parameter value far from wall (hence the level of width determination).

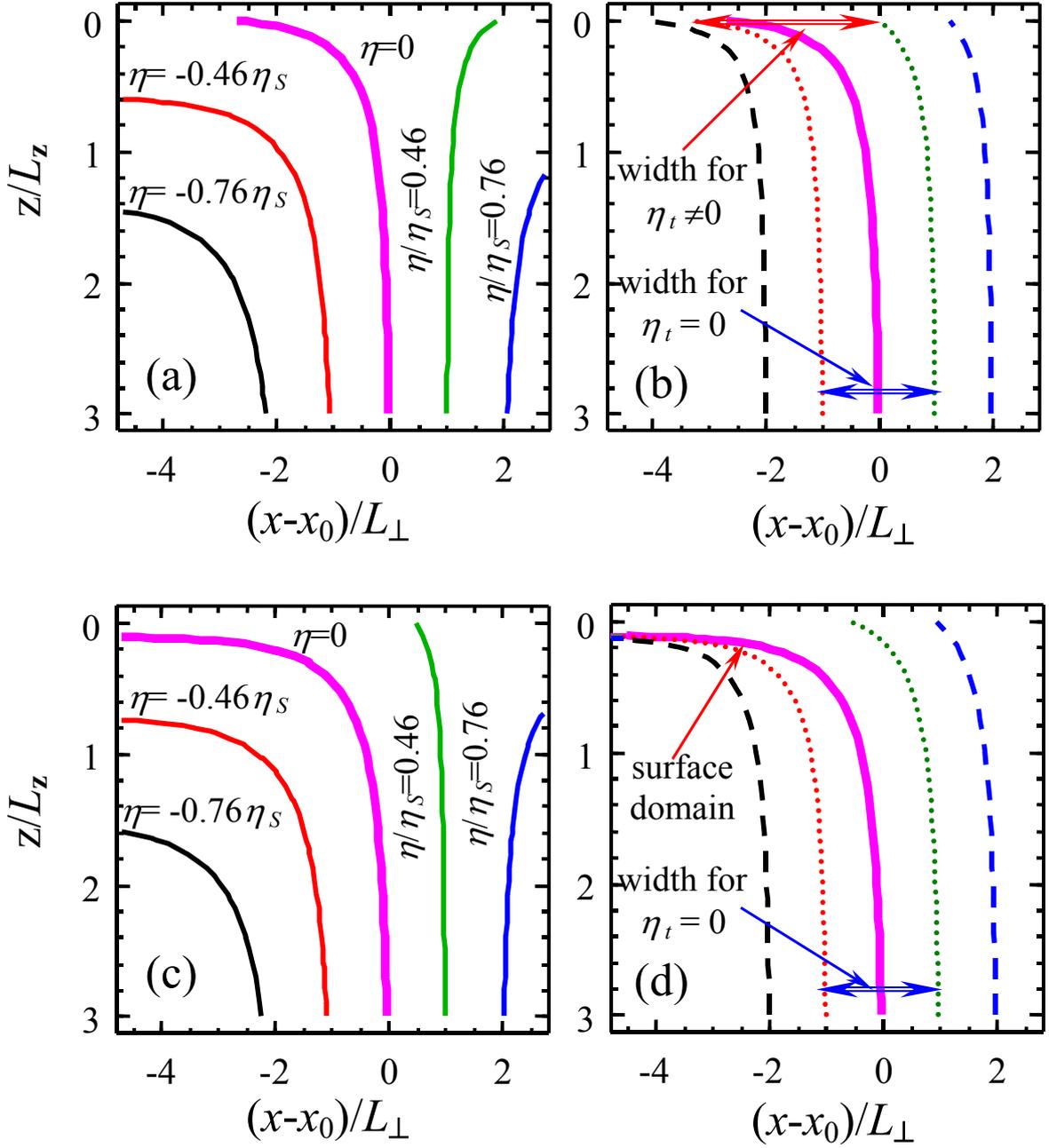

**Fig. 3**. (Color online) Schematics of the order parameter spatial distribution for $\eta_t=0.35\eta_S$ (panels (a) and (b)) and $\eta_t=0.55\eta_S$ (panels (c) and (d)). Left column (panels (a) and (c)) represents map contours of order parameter constant values marked near curves; the curves of fixed width at levels $\zeta=\pm0.76$ (dashed curves) and $\pm0.46$ (dotted curves) right column (panels (b) and (d)).

In our case the order parameter depends on z coordinate even far from the wall (and so does the level), thus width cannot be simply deduced from the contour maps like Figs. 3a and 3c. So, in order to illustrate the depth dependence of the wall width on the surface field, we calculated the curves, at which order parameters is equal to fixed ratio $\zeta$ (level) of order parameter far from wall but on the same depth, $\eta(x,z)=\pm\xi\eta(x\to\pm\infty,z)$ (see Fig. 3b and 3d).

Levels about 0.46 and 0.76 are chosen because $\tanh(0.5)\approx0.46$ and $\tanh(1)\approx0.76$, so that the width on the level 0.46 is very close to the value $2L_\perp$ for bulk walls.

It is seen from Figs. 3b and 3d, that wall width at the surface increases with $\eta_t$ increase, and formally reaches infinity when near-surface region transforms in single-domain state.

It is clear that the smaller extrapolation length $\lambda$ and/or the higher is the ratio $|\eta_t/\eta_S|$, the stronger is the bending effect. Under the absence of built-in surface field, the domain wall bending or broadening *is absent* as anticipated in linear approximation.

Calculated domain wall bending is caused by nonzero surface field of constant sign. Actually, built-in surface field distribution may be random (cf. discussion in Ref. 15). Some antisymmetric distribution of field like $\eta_t \sim \sigma \sim (x-x_0)$ should lead to the pure broadening effect. Also coordinate dependent surface field with variable polarity could lead to the wall pining in weak external field, since surface field would either promote nucleation of new domains,[15] or, as shown above, lead to the formation of surface domains in the regions with opposite value of order parameter. Thus external field will "push" the walls to the positions where $\eta_t \sim \sigma$ is close to zero. However, in weak external field asymmetric surface field would create restoring force preventing further displacement of the walls.

For typical material parameters $L_z$ values is about several lattice constants for ferroelectrics and much higher (up to hundred lattice constants) for magnetics. Scanning probe methods like PFM or SNDM measure *effective domain wall width* in ferroelectrics, since probe field penetrates in the depth of the sample at distances 1-10 nm, and so could recognize bended walls as broadened ones. Thus, built-in surface field is one of the possible mechanisms of domain wall broadening near the ferroic surfaces.

**Supplementary material**

Looking for the solution of Eqs.(3) in the form $p(x,z) = \int_{k \geq 0} dk\, q(k,x) \exp(-kz)$, one obtains the equations for the spectrum $q(k,x)$:

$$\left(-2\alpha + 3\alpha \operatorname{sech}^2\left(\frac{x-x_0}{2L_\perp}\right) + \gamma k^2\right) q(k,x) - \xi \frac{d^2}{dx^2} q(k,x) = 0 \,, \qquad \text{(S.1a)}$$

$$\int_{k \geq 0} dk(1+\lambda k) q(k,x) = \eta_t(x) - \eta_S \tanh\left(\frac{x-x_0}{2L_\perp}\right). \qquad \text{(S.1b)}$$

Eq.(4a) is linear inhomogeneous equation with x-dependent coefficient. The solution of homogeneous equation for $q(k,x)$ was derived as proposed in Ref.[28], namely:

$$q(k,x) = A(k) f(k, x-x_0) + B(k) f(k, x_0 - x), \qquad \text{(S.2a)}$$

$$f(k,x) = \exp\left(i\kappa(k)\frac{x}{2L_\perp}\right)\left(\kappa^2(k) - 2 + 3\operatorname{sech}^2\left(\frac{x}{2L_\perp}\right) + 3i\kappa(k)\tanh\left(\frac{x}{2L_\perp}\right)\right). \qquad \text{(S.2b)}$$

Where $A(k)$ and $B(k)$ are arbitrary functions of $k$, dispersion law $\kappa(k) = 2\sqrt{\frac{L_\perp^2}{\xi} k^2 \gamma - 1}$, while $k \geq \sqrt{\xi/\gamma} \cdot L_\perp^{-1}$. Then one obtains integral equation for $A(k)$ and $B(k)$ from the boundary condition (S.1b) as:

$$\int_{k \geq 0} dk(1+\lambda k)(A(k) f(k, x-x_0) + B(k) f(k, x_0 - x)) = \eta_t(x) - \eta_S \tanh\left(\frac{x-x_0}{2L_\perp}\right), \qquad \text{(S.3)}$$

Eq.(S.3) should be valid at arbitrary $x$. It reduces to the two Fredholm equations of the second order. Only numerical solutions are available.

Without built-in field, $\eta_t = \sigma = 0$, one obtains from Eq.(S.3) that $A(k) = -B(k) = -\eta_S \delta(k - k_0)/(6i\kappa(k))$, $\delta(k)$ is Dirac-delta function and $k_0 = \sqrt{\xi/\gamma} \cdot L_\perp^{-1}$. Finally linearized solution acquires the simplest form:

$$\eta(x,z) = \eta_S\left(1 - \frac{\exp(-z/L_z)}{1 + \lambda/L_z}\right)\tanh\left(\frac{x - x_0}{2L_\perp}\right), \quad \text{where} \quad L_z = \sqrt{-\gamma/2\alpha}. \quad (S.4)$$

As anticipated, for the absence of built-in surface field ($\eta_t = 0$), the domain wall bending or broadening *is absent* in linear approximation.